# IMPACT OF AN OBLIQUE BREAKING WAVE ON A WALL

**Jian-Jun SHU**
**School of Mechanical & Aerospace Engineering, Nanyang Technological University,**
**50 Nanyang Avenue, Singapore 639798**
**E-mail: mjjshu@ntu.edu.sg**

## Abstract

The purpose of this paper is to study the impact force of an oblique angled slamming wave acting on a rigid wall. In the present study the analytical approach is pursued based on a technique proposed by the author. A nonlinear theory in the context of potential flow is presented for determining accurately the free-surface profiles immediately after an oblique breaking wave impingement on the rigid vertical wall that suddenly starts from rest. The small-time expansion is taken as far as necessary to include the accelerating effect. The analytical solutions for the free-surface elevation are derived up to the third order. The results derived in this paper are of particular interest to the marine and offshore engineering industries, which will find the information useful for the design of ships, coastal and offshore.

## Introduction

One of the most devastating forces of nature is that of breaking waves. The destructive force of breaking waves is economically and physically detrimental and fatal. Hence, a considerable amount of research has been devoted to the study of the impact forces of breaking waves, particularly that of breaking waves impacting on a rigid wall, which is suddenly started from rest and made to move towards a fluid taper. Such studies can yield useful results that benefit designers of dams, ships, oil rigs, and other coastal and offshore structures that are directly exposed to the impact forces of breaking waves.

When a breaking wave strikes a wall, the impact produced is of short duration but considerable intensity. This direct collision produces impulsive pressure on the wall, similar to the problem of water impact in the initial stage; however, due to the highly nonlinear and transient nature of the problem, existing wave theories based on the assumption of small and finite amplitudes cannot accurately model the breaking wave force on the wall.

In reviewing the previous studies, one of the most important and unresolved questions is how the initial stage of the breaking wave impingement on the wall can be properly characterized and simulated. Cumberbatch [1] considered the case of symmetric normal impact of a water wedge on a wall, and Zhang *et al.* [2] took it further by studying an oblique impact. These two works assumed a prescribed function of the free surface profile near the wall: in Cumberbatch [1], a linear function was assumed, while in Zhang *et al.* [2], an exponential function was assumed. These two works were stemmed from an *ad hoc* assumption on the free-surface profiles close to the wall.

In Shu [3], an analytical approach was taken to solve the breaking wave problem for a normal wave without prescribed functions. It has been found that the free-surface profile close to wall is neither linear in Cumberbatch's assumption [1] nor exponential in Zhang *et al.*'s assumption [2]. This paper aims to follow the same analytical approach, but instead of a normal wave, we will derive and address the impact problem caused by an oblique angled wave.

In the present study, we do not assume any prescribed functions for the free surface profiles.

Since inertia forces dominate during small-time impacts, the effects of gravity, viscosity, and surface tension are neglected. The essential mechanism involved in the impact process can be described by the theoretical treatment of potential flow. A small portion of the breaker tip is initially cut off to produce a finite wetted area on the wall and a high spike in the consequent impact results from an acceleration of water towards the wall. We are interested in the brief time successive triggering of the nonlinear effects using a small-time expansion of the full, nonlinear initial/boundary value problem. The leading small-time expansion is taken to include the accelerating effect.

## Governing equations

We consider a rigid horizontal wall, being suddenly started from rest and made to move vertically with constant acceleration $a_1 = a_0 \cos(\beta\pi)$ towards a two-dimensional fluid taper with semi-angle $\alpha\pi\,(0 < \alpha < 1/2)$. A definition sketch of the flow is shown in Fig. 1. The axis of the fluid taper is at an angle $\beta\pi\,(0 < \beta < 1/2 - \alpha)$ to the vertical. Let us nondimensionalize time $t$ by $(L_1/a_1)^{1/2}$, distance $(x, y)$ by $L_1$, velocity $(u, v)$ by $(a_1 L_1)^{1/2}$, pressure $p$ by $\rho a_1 L_1$, where $L_1$ is the right-side wetted wall semi-length when the breaking wave just touches the wall at time $t = 0$ and $\rho$ is the density of the fluid. A mathematical statement of the above problem can now be written as

$$\frac{\partial u}{\partial x} + \frac{\partial v}{\partial y} = 0, \tag{1}$$

$$\frac{\partial u}{\partial t} + u\frac{\partial u}{\partial x} + v\frac{\partial u}{\partial y} = -\frac{\partial p}{\partial x}, \tag{2}$$

$$\frac{\partial v}{\partial t} + u\frac{\partial v}{\partial x} + v\frac{\partial v}{\partial y} = -\frac{\partial p}{\partial y}. \tag{3}$$

For negative time $t < 0$ everything is at rest,

$$u = v = 0,\ \eta_1 = 0,\ \eta_2 = 0 \ \text{ for } t < 0, \tag{4}$$

where $\eta_1$ and $\eta_2$ are the free surface ``elevations'' in the $x$ direction beyond the undisturbed surface. On the surface, kinematic and dynamic boundary conditions require

$$u = \frac{\partial \eta_1}{\partial t} + v\frac{\partial \eta_1}{\partial y},\ p = 0 \quad \text{on } x = 1 + y\tan[(\alpha - \beta)\pi] + \eta_1(y,t)$$

$$u = \frac{\partial \eta_2}{\partial t} + v\frac{\partial \eta_2}{\partial y},\ p = 0 \quad \text{on } x = -L_2 - y\tan[(\alpha + \beta)\pi] - \eta_2(y,t) \tag{5}$$

where the left-side wetted wall semi-length $L_2$ can be expressed in angles $\alpha\pi$ and $\beta\pi$, as follows:

$$L_2 = \frac{\cos[(\beta - \alpha)\pi]}{\cos[(\beta + \alpha)\pi]}. \tag{6}$$

On the wall surfaces, the normal velocity of fluid particles must be the same as that of the wall at all time

$$v = a_1 t \ \text{ on } \ y = a_1 t^2/2. \tag{7}$$

The pressure vanishes at infinity,

$$p \to 0 \ \text{ as } \ y \to \infty. \tag{8}$$

## Mathematical analysis

The full nonlinear initial/boundary value problem consists of equations (1)-(3) with conditions (4)-(8). These equations are solved analytically by employing a small-time expansion. We assume that

$$u(x, y, t) = u_1(x, y)t + O(t^2),$$
$$v(x, y, t) = v_1(x, y)t + O(t^2), \tag{9}$$

$$p(x, y, t) = p_0(x, y) + O(t), \tag{10}$$

$$\eta_1(y,t) = \eta_{12}(y)t^2 + O(t^3),$$
$$\eta_2(y,t) = \eta_{22}(y)t^2 + O(t^3). \tag{11}$$

The leading-order equations are

$$\frac{\partial u_1}{\partial x} + \frac{\partial v_1}{\partial y} = 0,\ u_1 = -\frac{\partial p_0}{\partial x},\ v_1 = -\frac{\partial p_0}{\partial y} \tag{12}$$

subject to the conditions

$$u_1 = 2\eta_{12},\ p_0 = 0 \quad \text{on } x = 1 + y\tan[(\alpha - \beta)\pi], \tag{13}$$

$$u_1 = 2\eta_{22}\,,\ \ p_0 = 0 \quad \text{on}\ \ x = -L_2 - y\tan[(\alpha+\beta)\pi], \tag{14}$$

$$v_1 = a_1 \ \ \text{on}\ \ y = 0, \tag{15}$$

$$p_0 \to 0 \ \ \text{as}\ \ y \to \infty. \tag{16}$$

It is clear that pressure $p_0$ satisfies the Laplace equation

$$\frac{\partial^2 p_0}{\partial x^2} + \frac{\partial^2 p_0}{\partial y^2} = 0. \tag{17}$$

Introducing a complex-conjugate function $q_0$ with respect to $p_0$, we can construct an analytic function

$$f_0(z) \equiv p_0 + iq_0\,,\ \ z = x + iy. \tag{18}$$

As shown in Fig. 2, the conformal mapping

$$z = \frac{(1+L_2)^{1-2\alpha}}{\mathrm{B}\left(1+\gamma_1; 1+\gamma_2; \dfrac{1}{1+L_2}\right)} \times \int_0^w (1-\tau)^{\gamma_1}(L_2+\tau)^{\gamma_2}\, d\tau - L_2 \tag{19}$$

given by the Schwarz-Christoffel transformation, maps the upper half of the $w$ plane $(w = \xi + i\zeta)$ onto the region occupied by the fluid. Here $\mathrm{B}$ is the incomplete Beta function defined by

$$\mathrm{B}(\nu;\mu;x) = \int_0^x \tau^{\nu-1}(1-\tau)^{\mu-1}\, d\tau, \tag{20}$$

$$\gamma_1 = \alpha - \beta - \frac{1}{2},\ \ \gamma_2 = \alpha + \beta - \frac{1}{2}. \tag{21}$$

Function $f_0$ is also analytic in the transformed variable $w$. On the free surfaces, which correspond to $\xi < 0$ and $\xi > 1$ on the positive real axis, $p_0$ vanishes. On the wall surface, which corresponds to the line segment $0 < \xi < 1$, we take $\partial p_0 / \partial n = a_1$. Therefore, along the real axis in the $w$ plane, we have

$$\mathrm{Re}(f_0) = 0 \ \ \text{on}\ \ -\infty < \xi < 0, \tag{22}$$

$$\mathrm{Re}\left(\frac{\partial f_0}{\partial n}\right) = a_1 \ \ \text{on}\ \ 0 < \xi < 1, \tag{23}$$

$$\mathrm{Re}(f_0) = 0 \ \ \text{on}\ \ 1 < \xi < \infty. \tag{24}$$

If $s(\xi)$ measures the distance from point **C** in Fig. 2 to any point on the wall surface, the Cauchy-Riemann equations give

$$\mathrm{Re}(f_0) = 0 \ \ \text{on}\ \ -\infty < \xi < 0, \tag{25}$$

$$\mathrm{Im}(f_0) = a_1 s(\xi) \ \ \text{on}\ \ 0 < \xi < 1, \tag{26}$$

$$\mathrm{Re}(f_0) = 0 \ \ \text{on}\ \ 1 < \xi < \infty, \tag{27}$$

where the distance $s(\xi)$ is given by (19) as

$$s(\xi) = \frac{(1+L_2)^{1-2\alpha}}{\mathrm{B}\left(1+\gamma_1; 1+\gamma_2; \dfrac{1}{1+L_2}\right)} \times \int_\xi^1 (1-\tau)^{\gamma_1}(L_2+\tau)^{\gamma_2}\, d\tau \ \ \text{on}\ \ 0 < \xi < 1. \tag{28}$$

If we introduce a new analytic function $g_0(w)$ by

$$g_0(w) = w^{-1/2}(1-w)^{-1/2} f_0(w), \tag{29}$$

the boundary conditions for $g_0(w)$ are unmixed

$$\mathrm{Im}(g_0) = 0 \ \ \text{on}\ \ -\infty < \xi < 0, \tag{30}$$

$$\mathrm{Im}(g_0) = a_1 \xi^{-1/2}(1-\xi)^{-1/2} s(\xi) \ \ \text{on}\ \ 0 < \xi < 1, \tag{31}$$

$$\mathrm{Im}(g_0) = 0 \ \ \text{on}\ \ 1 < \xi < \infty. \tag{32}$$

The analytic function $g_0(w)$ that is regular in the upper half $w$ plane and vanishes at infinity can be obtained from the Schwarz integral formula

$$g_0(w) = \frac{1}{\pi}\int_{-\infty}^{\infty} \frac{\mathrm{Im}(g_0)}{\tau - w}\, d\tau. \tag{33}$$

Substituting (29) -- (32) into (33), we have

$$f_0(w) = \frac{a_1 w^{1/2}(1-w)^{1/2}}{\pi} \times \int_0^1 \frac{s(\tau)}{\tau^{1/2}(1-\tau)^{1/2}(\tau - w)} d\tau . \quad (34)$$

From boundary conditions (13) to (16), we have

$$\left. \frac{\mathrm{Im}(\partial f_0 / \partial w)}{\mathrm{Im}(\partial z / \partial w)} \right|_{\zeta=0} = 2\eta_{12}(\xi) \text{ on } \xi > 1, \quad (35)$$

$$\left. \frac{\mathrm{Im}(\partial f_0 / \partial w)}{\mathrm{Im}(\partial z / \partial w)} \right|_{\zeta=0} = 2\eta_{22}(\xi) \text{ on } \xi < 0. \quad (36)$$

After some mathematical manipulation, we obtain

$$\eta_{12}(\xi) = \frac{a_1}{2(\xi - 1)\sin(\gamma_1 \pi)} \times \mathrm{B}(1+\gamma_1; 1+\gamma_2; \frac{1}{1+L_2}) \times \mathrm{B}(1+\gamma_1; 1+\gamma_2; \frac{1+L_2}{\xi + L_2}) \text{ on } \xi > 1, \quad (37)$$

$$\eta_{22}(\xi) = \frac{a_1}{2(L_2 - \xi)\sin(\gamma_2 \pi)} \times \mathrm{B}(1+\gamma_2; 1+\gamma_1; \frac{1}{1+L_2}) \times \mathrm{B}(1+\gamma_2; 1+\gamma_1; \frac{1+L_2}{1-\xi}) \text{ on } \xi < 0, \quad (38)$$

Impact free-surface profiles for different $\alpha$ and $\beta$ are shown in Figs. 3 and 4. It has been found that the free-surface profiles $\eta_{12}(y)$ and $\eta_{22}(y)$ close to the wall are proportional to $y^{-1/\gamma_1}$ and $y^{-1/\gamma_2}$, respectively, which are neither linear in Cumberbatch's assumption [1] nor exponential in Zhang *et al.*'s assumption [2].

## Conclusions

The problems of oblique breaking wave impingement on the wall and the free-surface profiles have been solved analytically by using a small-time expansion. The results obtained show that the free-surface profiles can be determined if the angles and acceleration of the oblique breaking wave are given. The results of this paper agree with the results of the case of a normal impact (Shu [3]) with angle $\beta = 0$. In addition, we have further confirmed that the free surface profile close to the wall is neither linear in Cumberbatch's assumption [1] nor exponential in Zhang *et al.*'s assumption [2].

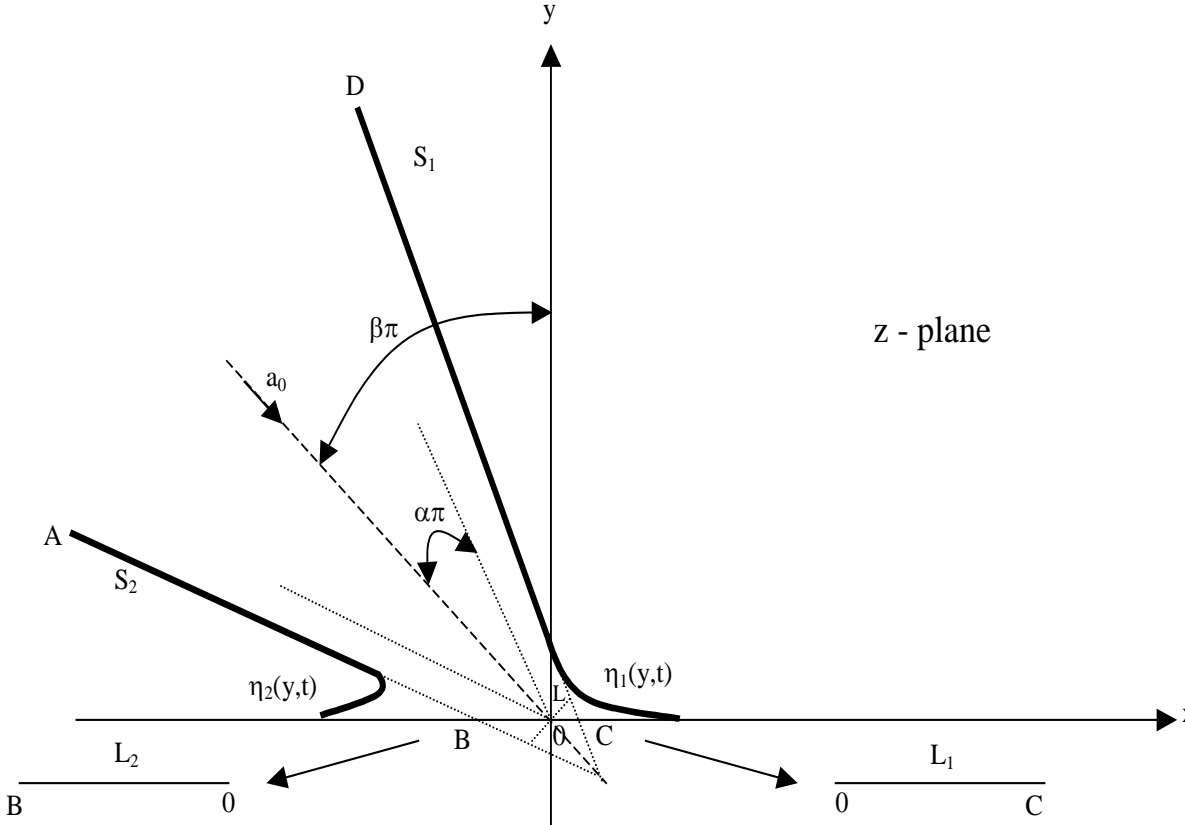

**Fig. 1: Fluid body in physical $z$ -plane**

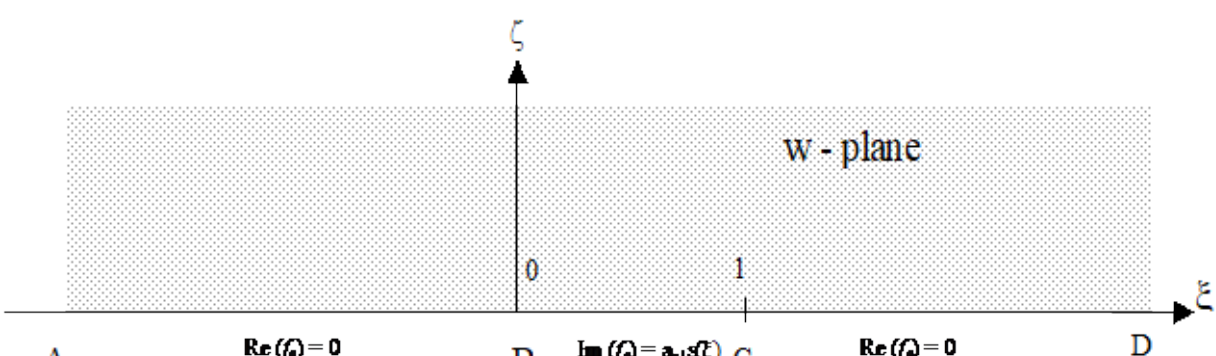

**Fig. 2: Physical $z$ plane conformally maps to upper half of $w$ plane.**

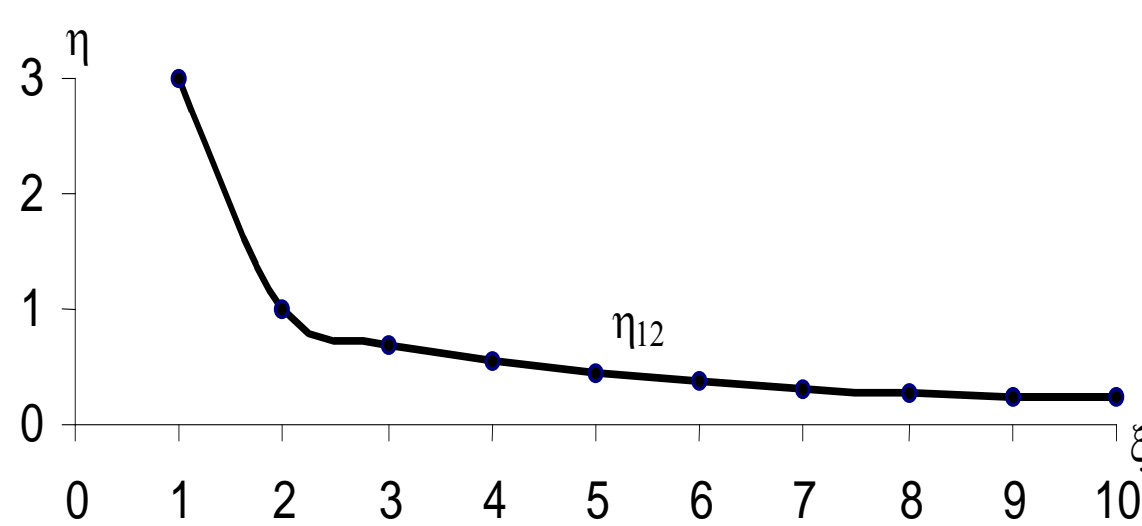

**Fig. 3: Impact free-surface profile $\eta_{12}(\xi)/a_1$ for $\alpha = 1/6$, $\beta = 0$**

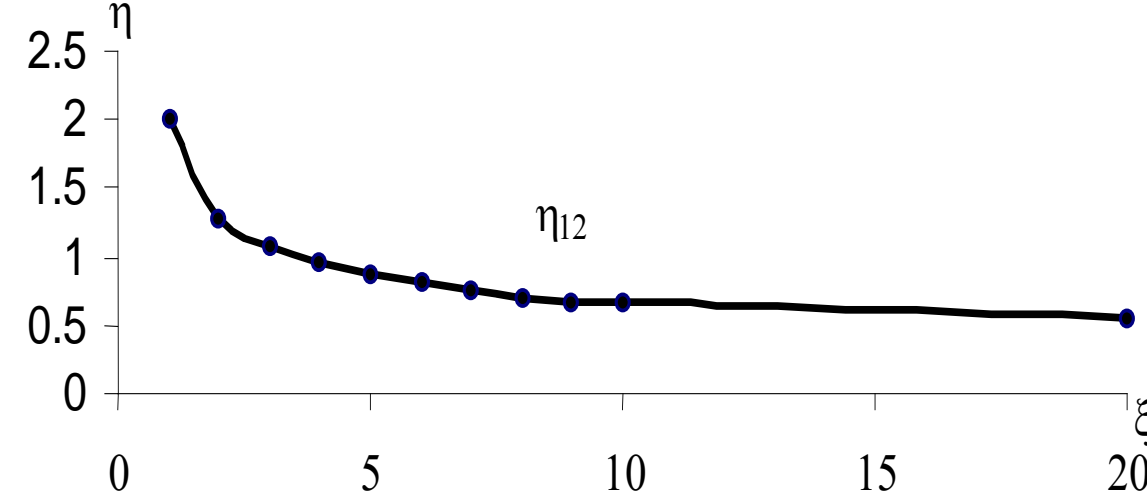

**Fig. 4: Impact free-surface profile $\eta_{12}(\xi)/a_1$ for** $\alpha = 1/6$, $\beta = 1/4$